# Acoustic force of the gravitational type

Ion Simaciu[1a], Gheorghe Dumitrescu[2b] and Zoltan Borsos[1]

[1] Petroleum-Gas University of Ploiești, Ploiești 100680, Romania

[2] High School Toma N. Socolescu, Ploiești, Romania

**Abstract:** In this paper it was shown that, under certain restrictive conditions, Bjerknes secondary forces are attractive and proportionate to the product of the virtual masses of the two bubbles.

## 1. Introduction

The secondary Bjerknes force is a mechanism of mutual interaction between bubbles oscillating in a sound field. Secondary Bjerknes forces manifest between pairs of bubbles in the same acoustic field. Under certain restrictive conditions, the secondary Bjerknes force can present a gravitational relation as we will further show in this paper. Our derivation leads to a form of the force independent on the sign of the expression of the force and proportional to the product of the cube of the two bubbles radii, $F_{Batr} \sim R_{01}^3 R_{02}^3$. This means that the secondary Bjerknes force is independent of the phase difference. We will show that this form also implies the proportionality with the product of the virtual masses (the added masses) of the oscillating bubble $m_i = 4\pi R_{0i}^3 \rho, i = 1, 2$ [1].

For two bubbles with different radii, secondary Bjerknes force is both attractive and repulsive. For identical bubbles these forces are attractive for any frequency up to a rank of the radius for which the forces become repulsive, i.e. when small variations of the two radii occurs [2, 3, 4]. We adopt this case in order to use it for the electromagnetic interaction where charged particles, e.g. electrons and positrons, have the same interaction cross section and therefore the same radius.

## 2. Setting the conditions for an attractive acoustic force that does not depend on the phase difference

In order to find the expression of the gravitational acoustic force between two bubbles, we will target to reach the following properties:
- be attractive, whereas the bubbles oscillating are in phase or in phase opposition (i.e. electrostatically, not dependent on the charges sign),
- be proportional to the product of the radii of bubbles (i.e. the product of the virtual masses of the two bubbles),

[a] isimaciu@yahoo.com, [b] meditatie@yahoo.com

- be dependent on the viscosity coefficient, $\beta_{vti} = \beta_{vi} + \beta_{thi}$, i.e. be a force emerging from the energy absorption from a plane wave which induces the oscillations of the bubbles and which is also converted into the thermal energy of the fluid and gas / vapor within the bubbles,
- have a significantly smaller value than that of the electrostatic acoustic force.

In order to infer the expression of gravitational force, we start from the expression of the Bjerknes secondary force [5]

$$F_B(r,\varphi) = -\frac{2\pi\rho\omega^2 R_{01}^3 R_{02}^3}{r^2} a_1 a_2 \cos\varphi \left[1 - \frac{a_1 a_2}{\cos\varphi} + \frac{a_1^2 + a_2^2}{4} + 2a_1 a_2 \cos\varphi + O(a^2, \cos\varphi)\right] =$$
$$-\frac{2\pi\rho\omega^2 R_{01}^3 R_{02}^3}{r^2} a_1 a_2 \left[\cos\varphi - a_1 a_2 + \frac{a_1^2 + a_2^2}{4}\cos\varphi + 2a_1 a_2 \cos^2\varphi + O(a^2, \cos\varphi)\right], \quad (1)$$

where

$$a_i = \frac{A}{\rho R_{0i}^2 \left[(\omega^2 - \omega_{0i}^2)^2 + 4\beta_i^2 \omega^2\right]^{1/2}}, \quad \varphi_i = \arctan\frac{2\beta_i \omega}{(\omega^2 - \omega_{0i}^2)}, i=1,2. \quad (2)$$

When bubbles oscillate in phase, $\varphi = \varphi_2 - \varphi_1 = 0$, and the expression of force (1) becomes

$$F_B(r,\varphi=0) = -\frac{2\pi\rho\omega^2 R_{01}^3 R_{02}^3}{r^2} a_1 a_2 \left[1 - a_1 a_2 + \frac{a_1^2 + a_2^2}{4} + 2a_1 a_2 + O(a^2)\right]. \quad (3)$$

When bubbles oscillate in phase opposition, $\varphi = \pi$, and the expression of force (1) becomes

$$F_B(r,\varphi=\pi) = -\frac{2\pi\rho\omega^2 R_{01}^3 R_{02}^3}{r^2} a_1 a_2 \left[-1 - a_1 a_2 - \frac{a_1^2 + a_2^2}{4} + 2a_1 a_2 + O(a^2)\right]. \quad (4)$$

One can see from equations (3) and (4) that the second and the fourth terms of equation (1) are independent of the sign of the phase function $\cos\varphi$. Then we adopt:

$$F_{Batr.}(r,\varphi) = -\frac{2\pi\rho\omega^2 R_{01}^3 R_{02}^3}{r^2} a_1 a_2 (2a_1 a_2 \cos^2\varphi - a_1 a_2) = -\frac{2\pi\rho\omega^2 R_{01}^3 R_{02}^3}{r^2} a_1^2 a_2^2 (2\cos^2\varphi - 1). \quad (5)$$

Continuing by subtracting in (5) the expression of the dimensional amplitude from (2), yields

$$F_{Batr.}(r) = -\frac{2\pi\omega^2 A^4}{r^2 \rho^3 R_{01} R_{02} \left[(\omega^2 - \omega_{01}^2)^2 + 4\beta_1^2 \omega^2\right]\left[(\omega^2 - \omega_{02}^2)^2 + 4\beta_2^2 \omega^2\right]} (2\cos^2\varphi - 1). \quad (6)$$

This force is attractive, regardless of the phase difference. This is a consequence of both scattering and absorption of plan wave energy by the bubble. This force is much smaller than the corresponding electroacoustic force (see (19) of [7]).

The gravitational acoustic force is proportional to the product of the masses and the absorption coefficients of both bubbles. We will adopt a damping coefficient which has three components:

$$\beta_i = \beta_{aci} + \beta_{vi} + \beta_{thi} = \frac{\omega^2 R_{0i}}{2u} + \frac{2\mu}{\rho R_{0i}^2} + \beta_{thi}, \quad (7)$$

an acoustic one ($(\omega^2 R_{0i})/(2u)$), one of the liquid viscosity ($\beta_{vi} = 2\mu/(\rho R_{0i}^2)$) and thermal one associated to the fluid viscosity of the bubbles gas or vapor ($\beta_{thi}$).

### 3. Thermal damping coefficient $\beta_{thi}$

According to the paper [6], the thermal damping constant ($2\beta_{thi} = \omega\delta_{thi}$) is

$$\beta_{thi} = \frac{3(\gamma-1)\left[X_i(\sinh X_i + \sin X_i) - 2(\cosh X_i - \cos X_i)\right]\omega_{0i}^2}{2X_i\left[X_i(\cosh X_i - \cos X_i) + 3(\gamma-1)(\sinh X_i - \sin X_i)\right]\omega}, \quad (8)$$

with $X_i = R_{0i}(2\omega/\chi)^{1/2}$, $\gamma$ the ratio of specific heats, $\chi = K/(\rho_g c_P) \cong \mu_g/(\rho_g \gamma)$ the thermal diffusivity of gas in bubbles and $K$ the thermal conductivity.

For $X_i = R_{0i}(2\omega/\chi)^{1/2} \ll 1$, the parentheses in (8) can be approached by

$$\sinh X_i + \sin X_i = 2X_i\left(1 + \frac{X_i^4}{120}\right), \sinh X_i - \sin X_i = \frac{X_i^3}{3}\left(1 + \frac{X_i^4}{840}\right), \cosh X_i - \cos X_i = X_i^2\left(1 + \frac{X_i^4}{360}\right), \quad (9)$$

since

$$\sinh X_i = \frac{e^{X_i} - e^{-X_i}}{2} \cong X_i + \frac{X_i^3}{6} + \frac{X_i^5}{120} + \frac{X_i^7}{5040} + ..., \quad \cosh X_i = \frac{e^{X_i} + e^{-X_i}}{2} \cong 1 + \frac{X_i^2}{2} + \frac{X_i^4}{24} + \frac{X_i^6}{720} + ...,$$

$$\sin X_i \cong X_i - \frac{X_i^3}{6} + \frac{X_i^5}{120} - \frac{X_i^7}{5040} + ..., \quad \cos X_i \cong 1 - \frac{X_i^2}{2} + \frac{X_i^4}{24} - \frac{X_i^6}{720} + ... \quad (10)$$

Subtracting (9) in (8), it follows

$$\beta_{thi} \cong \frac{(\gamma-1)X_i^2 \omega_{0i}^2}{60\gamma\omega} = \frac{(\gamma-1)R_{0i}^2 \omega_{0i}^2}{30\gamma\chi}. \quad (11)$$

From the same paper [6], it is shown that the natural angular frequency (the resonance frequency), with $p_{eff} = 3\gamma p_0$, is

$$\omega_{0i} = \left(\frac{p_{eff}}{\rho R_{0i}^2}\right)^{1/2}\left[1 + \frac{3(\gamma-1)(\sinh X_i - \sin X_i)}{X_i(\cosh X_i - \cos X_i)}\right]^{-1/2} \cong \left(\frac{p_{eff}}{\rho\gamma R_{0i}^2}\right)^{1/2}. \quad (12)$$

Considering the pressure caused by the surface tension [5], the natural angular frequency becomes

$$\omega_{0i} \cong \left[3\gamma\left(\frac{p_0}{\rho R_{0i}^2} + \frac{2\sigma}{\rho R_{0i}^3}\right) - \frac{2\sigma}{\rho R_{0i}^3}\right]^{1/2}\frac{1}{\gamma^{1/2}} = \left(\frac{p_{eff}(\sigma)}{\rho\gamma R_{0i}^2}\right)^{1/2}. \quad (13)$$

Subtracting (13) into (11), one can approach

$$\beta_{thi} \cong \frac{(\gamma-1)R_{0i}^2\omega_{0i}^2}{30\gamma\chi} = \frac{(\gamma-1)p_{eff}(\sigma)}{30\gamma^2\rho\chi}. \tag{14a}$$

If $3\gamma p_0 \gg (3\gamma-1)2\sigma/R_{0i}$, it results $p_{eff} \cong 3\gamma p_0$ and the thermal damping coefficient is independent of the bubble radius

$$\beta_{thi} \cong \frac{(\gamma-1)R_{0i}^2\omega_{0i}^2}{30\gamma\chi} = \frac{(\gamma-1)p_0}{10\gamma\rho\chi}. \tag{14b}$$

If $2(3\gamma-1)\sigma/R_{0i} \gg 3\gamma p_0$, it results $p_{eff} \cong 2(3\gamma-1)\sigma/R_{0i}$ and the thermal damping coefficient is dependent of the bubble radius

$$\beta_{thi} \cong \frac{(3\gamma-1)(\gamma-1)\sigma}{15\gamma^2\rho\chi R_{0i}}. \tag{14c}$$

For $X_i = R_{0i}(2\omega/\chi)^{1/2} \gg 1$, the parentheses in (8) can be approached by

$$\sinh X_i + \sin X_i \cong \frac{e^{X_i}}{2}, \sinh X_i - \sin X_i \cong \frac{e^{X_i}}{2}, \cosh X_i - \cos X_i \cong \frac{e^{X_i}}{2}, \tag{15}$$

since

$$\sinh X_i = \frac{e^{X_i}-e^{-X_i}}{2} \cong \frac{e^{X_i}}{2}, \quad \cosh X_i = \frac{e^{X_i}+e^{-X_i}}{2} \cong \frac{e^{X_i}}{2}, \quad -1 \le \sin X_i \le 1, \quad -1 \le \cos X_i \le 1. \tag{16}$$

Hence, in this case, subtracting (15) into (8), the thermal damping coefficient becomes

$$\beta_{thi} \cong \frac{3(\gamma-1)\omega_{0i}^2}{2X_i\omega} = \frac{3(\gamma-1)\omega_{0i}^2\chi^{1/2}}{2\omega\sqrt{2\omega}R_{0i}} = \frac{3(\gamma-1)p_{eff}\chi^{1/2}}{2\gamma\omega\sqrt{2\omega}\rho R_{0i}^3}. \tag{17a}$$

If $3\gamma p_0 \gg 2(3\gamma-1)\sigma/R_{0i}$, it results $p_{eff} \cong 3\gamma p_0$ and the thermal damping coefficient is dependent of the bubble radius

$$\beta_{thi} = \frac{3(\gamma-1)p_{eff}\chi^{1/2}}{2\gamma\omega\sqrt{2\omega}\rho R_{0i}^3} = \frac{9(\gamma-1)\chi^{1/2}p_0}{2\omega\sqrt{2\omega}\rho R_{0i}^3}. \tag{17b}$$

If $2(3\gamma-1)\sigma/R_{0i} \gg 3\gamma p_0$, it results $p_{eff} \cong 2(3\gamma-1)\sigma/R_{0i}$ and the thermal damping coefficient

$$\beta_{thi} = \frac{3(\gamma-1)p_{eff}\chi^{1/2}}{2\gamma\omega\sqrt{2\omega}\rho R_{0i}^3} = \frac{3(3\gamma-1)(\gamma-1)\chi^{1/2}\sigma}{\gamma\omega\sqrt{2\omega}\rho R_{0i}^4} \tag{17c}$$

is dependent of the bubble radius.

## 4. Requirements for the force of the gravitational attractive type

### 4.1. Attractive force at resonance

When we address the case of identical bubbles, the force (6) can be rewritten using (7) as

$$F_{Batr}(r) = \frac{-2\pi\omega^2 A^4}{r^2 \rho^3 R_0^2 \left[ \left(\omega^2 - \omega_0^2\right)^2 + 4\left(\frac{\omega^2 R_0}{2u} + \frac{2\mu}{\rho R_0^2} + \beta_{th}\right)^2 \omega^2 \right]^2}. \quad (18)$$

Depending on the order of the scale adopted for $\omega, \omega_{0i}, \beta_{aci}, \beta_{vi}$ and $\beta_{thi}$, the interaction can be influenced by the scattering of the plane wave or by the absorption of the energy of the plane wave and conversion to thermal energy.

At resonance [5], $\omega^2 \cong \omega_0^2$, the relationship (18) becomes

$$F_{Batrr}(r) = \frac{-\pi A^4}{8 r^2 \rho^3 R_0^2 \left( \frac{\omega_0^2 R_0}{2u} + \frac{2\mu}{\rho R_0^2} + \beta_{th0} \right)^4 \omega_0^2}. \quad (19)$$

For $\beta_{ac0} \gg \beta_{vt0}$, the expression of the attractive force becomes

$$F_{Batrr}(r) \cong \frac{-2\pi u^4 A^4}{r^2 \rho^3 R_0^6 \omega_0^{10}} \left(1 - \frac{4u\mu}{\rho\omega_0^2 R_0^3} - \frac{2u\beta_{th0}}{\omega_0^2 R_0}\right)^4 = \frac{-2\pi\gamma^5 R_0^4 A^4 \rho^2 u^4}{r^2 p_{eff}^5} \left(1 - \frac{4\gamma u\mu}{p_{eff} R_0} - \frac{2\gamma u \rho R_0 \beta_{th0}}{p_{eff}}\right)^4 =$$

$$\frac{-2\pi\gamma^5 R_0^4 A^4 \rho^2 u^4}{r^2 p_{eff}^5} \left(1 - \frac{8\gamma u \mu}{p_{eff} R_0} - \frac{4\gamma u \rho R_0 \beta_{th0}}{p_{eff}} + \frac{16\gamma^2 u^2 \rho\mu\beta_{th0}}{p_{eff}^2} + \frac{16\gamma^2 u^2 \mu^2}{p_{eff}^2 R_0^2} + \frac{4\gamma^2 \rho^2 u^2 R_0^2 \beta_{th0}^2}{p_{eff}^2}\right)^2 =$$

$$\frac{-2\pi\gamma^5 R_0^4 A^4 \rho^2 u^4}{r^2 p_{eff}^5} \left(1 + \frac{2^5 \gamma^2 3 u^2 \mu^2}{p_{eff}^2 R_0^2} + \frac{2^3 3 \gamma^2 \rho^2 u^2 R_0^2 \beta_{th0}^2}{p_{eff}^2} + \frac{2^7 3 \gamma^4 \rho^2 u^4 \mu^2 \beta_{th0}^2}{p_{eff}^4} + \frac{2^8 \gamma^4 u^4 \mu^4}{p_{eff}^4 R_0^4} + \right. \quad (20a)$$

$$\left. \frac{2^4 \gamma^4 \rho^4 u^4 R_0^4 \beta_{th0}^4}{p_{eff}^4} + \frac{2^5 3 \gamma^2 u^2 \rho\mu\beta_{th0}}{p_{eff}^2} + \frac{2^9 \gamma^4 \rho u^4 \mu^3 \beta_{th0}}{p_{eff}^4 R_0^2} + \frac{2^7 \gamma^4 \rho^3 u^4 \mu R_0^2 \beta_{th0}^3}{p_{eff}^4}\right).$$

If the thermal damping coefficient is in accordance with the relationship (14a), the expression (20a) of the attractive force becomes

$$F_{Batrr}(r) \cong \frac{-2\pi\gamma^5 R_0^4 A^4 \rho^2 u^4}{r^2 p_{eff}^5} \left(1 + \frac{2^5 \gamma^2 3 u^2 \mu^2}{p_{eff}^2 R_0^2} + \frac{2^3 3 \gamma^2 \rho^2 u^2 R_0^2 \beta_{th0}^2}{p_{eff}^2} + \frac{2^7 3 \gamma^4 \rho^2 u^4 \mu^2 \beta_{th0}^2}{p_{eff}^4} + \frac{2^8 \gamma^4 u^4 \mu^4}{p_{eff}^4 R_0^4} + \right.$$

$$\left. \frac{2^4 \gamma^4 \rho^4 u^4 R_0^4 \beta_{th0}^4}{p_{eff}^4} + \frac{2^5 3 \gamma^2 u^2 \rho\mu\beta_{th0}}{p_{eff}^2} + \frac{2^9 \gamma^4 \rho u^4 \mu^3 \beta_{th0}}{p_{eff}^4 R_0^2} + \frac{2^7 \gamma^4 \rho^3 u^4 \mu R_0^2 \beta_{th0}^3}{p_{eff}^4}\right) =$$

$$\frac{-2\pi\gamma^5 R_0^4 A^4 \rho^2 u^4}{r^2 p_{eff}^5} \left(1 + \frac{2^5 3 \gamma^2 u^2 \mu^2}{p_{eff}^2 R_0^2} + \frac{2(\gamma-1)^2 u^2 R_0^2}{5^2 3 \gamma^2 \chi^2} + \frac{2^5 (\gamma-1)^2 u^4 \mu^2}{5^2 3 p_{eff}^2 \chi^2} + \frac{2^8 \gamma^4 u^4 \mu^4}{p_{eff}^4 R_0^4} + \right. \quad (20b)$$

$$\left. \frac{2(\gamma-1)^4 u^4 R_0^4}{3^4 5^4 \gamma^4 \chi^4} + \frac{2^4 (\gamma-1) u^2 \mu}{5 \chi p_{eff}} + \frac{2^8 \gamma^2 (\gamma-1) u^4 \mu^3}{15 \chi p_{eff}^3 R_0^2} + \frac{2^4 (\gamma-1)^3 u^4 \mu R_0^2}{3^3 5^3 \gamma^2 \chi^3 p_{eff}}\right).$$

If the thermal damping coefficient is in accordance with the relationship (17a), $\beta_{th0} = 3(\gamma-1) p_{eff} \chi^{1/2} / [2\gamma\omega_0 \sqrt{2\omega_0} \rho R_{0i}^3] = 3(\gamma-1) p_{eff}^{1/4} \chi^{1/2} / (2^{3/2} \gamma \rho^{1/4} R_{0i}^{3/2})$, the expression (20a) of the attractive force becomes

$$F_{Batrr}(r) \cong \frac{-2\pi\gamma^5 R_0^4 A^4 \rho^2 u^4}{r^2 p_{eff}^5} \left( 1 + \frac{2^5 \gamma^2 3 u^2 \mu^2}{p_{eff}^2 R_0^2} + \frac{2^3 3 \gamma^2 \rho^2 u^2 R_0^2 \beta_{th0}^2}{p_{eff}^2} + \frac{2^7 3 \gamma^4 \rho^2 u^4 \mu^2 \beta_{th0}^2}{p_{eff}^4} + \frac{2^8 \gamma^4 u^4 \mu^4}{p_{eff}^4 R_0^4} + \right.$$

$$\left. \frac{2^4 \gamma^4 \rho^4 u^4 R_0^4 \beta_{th0}^4}{p_{eff}^4} + \frac{2^5 3 \gamma^2 u^2 \rho \mu \beta_{th0}}{p_{eff}^2} + \frac{2^9 \gamma^4 \rho u^4 \mu^3 \beta_{th0}}{p_{eff}^4 R_0^2} + \frac{2^7 \gamma^4 \rho^3 u^4 \mu R_0^2 \beta_{th0}^3}{p_{eff}^4} \right) =$$

$$\frac{-2\pi\gamma^5 R_0^4 A^4 \rho^2 u^4}{r^2 p_{eff}^5} \left( 1 + \frac{2^5 3 \gamma^2 u^2 \mu^2}{p_{eff}^2 R_0^2} + \frac{3^3 (\gamma-1)^2 \chi \rho^{3/2} u^2}{p_{eff}^{3/2} R_0} + \frac{2^4 3^3 \gamma^2 (\gamma-1)^2 \rho^{3/2} u^4 \mu^2 \chi}{p_{eff}^{7/2} R_0^3} + \right. \quad (20c)$$

$$\frac{2^8 \gamma^4 u^4 \mu^4}{p_{eff}^4 R_0^4} + \frac{3^4 \gamma^4 (\gamma-1)^4 \rho^3 u^4 \chi^2}{2^2 \gamma^4 p_{eff}^3 R_0^2} + \frac{2^{7/2} 3^2 \gamma (\gamma-1) u^2 \rho^{3/4} \mu \chi^{1/2}}{p_{eff}^{7/4} R_0^{3/2}} +$$

$$\left. \frac{2^{15/2} 3 \gamma^3 (\gamma-1) \rho^{3/4} u^4 \mu^3 \chi^{1/2}}{p_{eff}^{15/4} R_0^{7/2}} + \frac{2^{5/2} 3^3 \gamma (\gamma-1)^3 \rho^{9/4} u^4 \mu \chi^{3/2}}{p_{eff}^{13/4} R_0^{5/2}} \right).$$

The third and ninth terms in (20b), for $p_{eff} \cong 3\gamma p_0$, are attractive forces which is proportional to the radius having exponent 6 and hence is related to the virtual mass $m = 4\pi R_0^3 \rho$, as the square of it

$$F_{Batrr}(R_0^6, r) \cong \frac{-2\pi R_0^4 A^4 \rho^2 u^4}{r^2 3^5 p_0^5} \left( \frac{2(\gamma-1)^2 u^2 R_0^2}{5^2 3 \gamma^2 \chi^2} + \frac{2^4 (\gamma-1)^3 u^4 \mu R_0^2}{3^4 5^3 \gamma^3 \chi^3 p_0} \right) =$$

$$\frac{-2^2 \pi (\gamma-1)^2 R_0^6 A^4 \rho^2 u^6}{r^2 3^6 5^2 \gamma^2 p_0^5 \chi^2} \left[ 1 + \frac{2^3 (\gamma-1) u^2 \mu}{3^3 5 \gamma p_0 \chi} \right] = \quad (21)$$

$$\frac{-m^2}{r^2} \left[ \frac{(\gamma-1)^2 A^4 u^6}{2^2 3^6 5^2 \pi \gamma^2 p_0^5 \chi^2} \right] \left[ 1 + \frac{2^3 (\gamma-1) u^2 \mu}{3^3 5 \gamma p_0 \chi} \right].$$

At resonance, the first term of the gravitational acoustic force (21) depends on the thermal diffusivity of gas/ vapor, $\chi \cong \mu_g/(\rho_g \gamma)$. It follows that this force is the effect of the wave energy absorption by the gas/vapor in the bubble. The second term also depends on the viscosity coefficient, $\mu$, of the liquid and so is the effect of the energy absorption by the liquid.

For $p_{eff} \cong 2(3\gamma-1)\sigma/R_0$, in the expression (20b) of the attractive force, there is no term proportional to the radius that has exponent 6.

In the expression (20c) of the attractive force, for $p_{eff} \cong 3\gamma p_0$ and $p_{eff} \cong 2(3\gamma-1)\sigma/R_0$, there is no term proportional to the radius that has exponent 6.

The attractive force (21) is proportional to the radius having the exponent 6, i.e. the square of the virtual masses of the bubbles in the oscillation motion [1]. The condition $X = R_0 (2\omega_0/\chi)^{1/2} << 1$ must also be fulfilled so that $\beta_{th}$, in accordance with the relationship (14b), to be independent of the radius.

When $\beta_\upsilon \gg \beta_{ac0}$, the expression of force (19) may be written as

$$F_{Batrr}(r) = \frac{-\pi A^4 \rho R_0^6}{2^7 r^2 \omega_0^2 \mu^4 \left(1 + \frac{\omega_0^2 R_0^3}{2^2 u\mu} + \frac{\rho R_0^2 \beta_{th0}}{2\mu}\right)^4} \cong \frac{-\pi \gamma A^4 \rho^2 R_0^8}{2^7 r^2 \mu^4 p_{eff}} \left(1 - \frac{R_0 p_{eff}}{2^2 \gamma u \mu} - \frac{\rho R_0^3 \beta_{th0}}{2\mu}\right)^4. \quad (22)$$

Regardless of the expressions (14) and (17) of the thermal damping coefficient at resonance, the force expression (22) has no terms proportional to the radius with the exponent 6.

If $\beta_{th} \gg \beta_{ac}$ and $\beta_{th} \gg \beta_\upsilon$, the expression (19) of force becomes

$$F_{Batrr}(r) = \frac{-\pi A^4}{8 r^2 \rho^3 R_0^2 \omega_0^2 \beta_{th0}^4 \left(1 + \frac{\omega_0^2 R_0}{2 u \beta_{th0}} + \frac{2\mu}{\rho R_0^2 \beta_{th0}}\right)^4} \cong \frac{-\pi \gamma A^4}{8 r^2 \rho^2 p_{eff} \beta_{th0}^4} \left(1 - \frac{p_{eff}}{2\gamma \rho u R_0 \beta_{th0}} - \frac{2\mu}{\rho R_0^2 \beta_{th0}}\right)^4. \quad (23)$$

Regardless of the expressions (14) and (17) of the thermal damping coefficient at resonance, the force expression (23) has no terms proportional to the radius with the exponent 6.

It follows that, at resonance, the attractive force fulfills the criteria listed in the second section for the case $\beta_{ac0} \gg \beta_{\upsilon 0t}$, according to the expression (21).

### 4.2. Expression of the attractive force for $\omega \ll \omega_0$ and $\omega \gg \omega_0$

When $\omega \ll \omega_0$, the square bracket from (18) can be approached as

$$\left[(\omega^2 - \omega_0^2)^2 + 4\beta^2 \omega^2\right] = \omega_0^4 \left[\left(\frac{\omega^2}{\omega_0^2} - 1\right)^2 + \frac{4\beta^2 \omega^2}{\omega_0^4}\right] \cong \omega_0^4 \left(1 - \frac{\omega^2}{\omega_0^2} + \frac{4\omega^2 \beta^2}{\omega_0^4}\right), \frac{\beta}{\omega_0} > \frac{\omega}{\omega_0}. \quad (24)$$

Subtracting (24) in (18), one can express the attractive force as

$$F_{Batr}(r) \cong \frac{-2\pi \omega^2 A^4}{r^2 \rho^3 R_0^2 \omega_0^8 \left(1 - \frac{\omega^2}{\omega_0^2} + \frac{4\omega^2 \beta^2}{\omega_0^4}\right)^2} = \frac{-2\pi \gamma^4 \omega^2 A^4 \rho R_0^6}{r^2 p_{eff}^4 \left(1 - \frac{\gamma \rho R_0^2 \omega^2}{p_{eff}} + \frac{4\gamma^2 \rho^2 R_0^4 \omega^2 \beta^2}{p_{eff}^2}\right)^2} \cong$$

$$\frac{-2\pi \gamma^4 \omega^2 A^4 \rho R_0^6}{r^2 p_{eff}^4} \left[1 + \frac{4\gamma \rho R_0^2 \omega^2}{p_{eff}} + 3\left(\frac{2\gamma \rho R_0^2 \omega \beta}{p_{eff}}\right)^4 + \ldots\right] \cong \quad (25a)$$

$$\frac{-2\pi \gamma^4 \omega^2 A^4 \rho R_0^6}{r^2 p_{eff}^4} \left[1 + 3\left(\frac{2\gamma \rho R_0^2 \omega \beta}{p_{eff}}\right)^4 + \ldots\right] = \frac{-2\pi \gamma^4 \omega^2 A^4 \rho R_0^6}{r^2 p_{eff}^4} \left[1 + 48\omega^4 \left(\frac{\gamma \rho R_0^2 \beta}{p_{eff}}\right)^4 + \ldots\right].$$

If $p_{eff} \cong 3\gamma p_0$, the expression (25a) of the attractive force becomes

$$F_{Batr}(r) \cong \frac{-2\pi \omega^2 A^4 \rho R_0^6}{3^4 p_0^4 r^2} \left[1 + 3\left(\frac{2\rho R_0^2 \omega \beta}{3 p_0}\right)^4 + \ldots\right] = \frac{-2\pi \omega^2 A^4 \rho R_0^6}{3^4 p_0^4 r^2} \left[1 + 48\omega^4 \left(\frac{\rho R_0^2 \beta}{3 p_0}\right)^4 + \ldots\right]. \quad (25b)$$

If $p_{eff} \cong 2(3\gamma - 1)\sigma/R_0$, the expression (25a) of the attractive force becomes

$$F_{Batr}(r) \cong \frac{-\pi\gamma^4\omega^2 A^4 \rho R_0^{10}}{2^3(3\gamma-1)^4 \sigma^4 r^2}\left[1+3\omega^4\left(\frac{\gamma\rho R_0^3 \beta}{(3\gamma-1)\sigma}\right)^4+...\right]. \tag{25c}$$

The first term in (25b) is an attractive force which is proportional to the radius having exponent 6 and hence is related to the virtual mass $m = 4\pi R_0^3 \rho$, as the square of it. The second is attractive and independent of the radius, unless the terms $\left(\rho R_0^2 \beta/(3p_0)\right)^4$

$$\left(\frac{\rho R_0^2 \beta}{3p_0}\right)^4 = \frac{\rho^4 R_0^8 (\beta_{ac}+\beta_\upsilon+\beta_{th})^4}{3^4 p_0^4} = \frac{\rho^4 R_0^8}{3^4 p_0^4}\left(\frac{\omega^2 R_0}{2u}+\frac{2\mu}{\rho R_0^2}+\beta_{th}\right)^4 \cong$$

$$\frac{\rho^4 R_0^8}{3^4 p_0^4}\left[\left(\frac{2\mu}{\rho R_0^2}\right)^4+\beta_{th}^4+4\left(\frac{2\mu}{\rho R_0^2}\right)^3\beta_{th}+6\left(\frac{2\mu}{\rho R_0^2}\right)^2\beta_{th}^2+4\left(\frac{2\mu}{\rho R_0^2}\right)\beta_{th}^3+4\left(\frac{\omega^2 R_0}{2u}\right)\beta_{th}^3+...\right] = \tag{26a}$$

$$\frac{2^4 \mu^4}{3^4 p_0^4}+\frac{\rho^4 R_0^8}{3^4 p_0^4}\beta_{th}^4+\frac{2^5 \mu^3 \rho R_0^2}{3^4 p_0^4}\beta_{th}+\frac{2^3 \mu^2 \rho^2 R_0^4}{3^3 p_0^4}\beta_{th}^2+\frac{2^3 \mu\rho^3 R_0^6}{3^4 p_0^4}\beta_{th}^3+\frac{2\rho^4 \omega^2 R_0^9}{3^4 up_0^4}\beta_{th}^3+...$$

is independent of the radius. In the expression (26a), only the first term and the sixth term, with $\beta_{th}$ having the expression (17b), are independent of the radius

$$\left(\frac{\rho R_0^2 \beta}{3p_0}\right)^4 \cong \frac{2^4 \mu^4}{3^4 p_0^4}+\frac{2\rho^4 \omega^2 R_0^9}{3^4 up_0^4}\beta_{th}^3+... = \frac{2^4 \mu^4}{3^4 p_0^4}+\frac{3^2(\gamma-1)^3 \rho\chi^{3/2}}{2^{7/2}\omega^{5/2}up_0}+... \tag{26b}$$

Under the above conditions, the expression of the attractive force (25b) becomes

$$F_{Batr}(r,R_0^6) \cong \frac{-2\pi\omega^2 A^4 \rho R_0^6}{3^4 p_0^4 r^2}\left[1+48\omega^4\left(\frac{\rho R_0^2 \beta}{3p_0}\right)^4+...\right] =$$

$$\frac{-2\pi\omega^2 A^4 \rho R_0^6}{3^4 p_0^4 r^2}\left[1+\frac{2^8 \mu^4 \omega^4}{3^3 p_0^4}+\frac{2^{1/2} 3^3 (\gamma-1)^3 \rho\omega^{3/2}\chi^{3/2}}{up_0}+...\right] = \tag{27}$$

$$\frac{-m^2}{r^2}\left[\frac{\omega^2 A^4}{2^3 3^4 \pi p_0^4 \rho}\right]\left[1+\frac{2^8 \mu^4 \omega^4}{3^3 p_0^4}+\frac{2^{1/2} 3^3 (\gamma-1)^3 \rho\omega^{3/2}\chi^{3/2}}{up_0}+...\right].$$

We have demonstrated that, for $\omega \ll \omega_0$, the gravitational acoustic forces are the effect of the induction wave scattering (first term) and the absorption of the energy of the wave by liquid and gas / vapor (the second and third terms).

The first term in (25c) is an attractive force which is proportional to the radius having exponent 10. The expression (25c) is an attractive force of gravitational type, only the terms $\left[\rho R_0^3 \beta/(3\gamma-1)\sigma\right]^4$

$$\left(\frac{\gamma\rho R_0^3 \beta}{(3\gamma-1)\sigma}\right)^4 = \frac{\gamma^4 \rho^4 R_0^{12}(\beta_{ac}+\beta_\upsilon+\beta_{th})^4}{(3\gamma-1)^4 \sigma^4} = \frac{\gamma^4 \rho^4 R_0^{12}}{(3\gamma-1)^4 \sigma^4}\left(\frac{\omega^2 R_0}{2u}+\frac{2\mu}{\rho R_0^2}+\beta_{th}\right)^4 \cong$$

$$\frac{\gamma^4 \rho^4 R_0^{12}}{(3\gamma-1)^4 \sigma^4}\left[\left(\frac{2\mu}{\rho R_0^2}\right)^4 + \beta_{th}^4 + 4\left(\frac{2\mu}{\rho R_0^2}\right)^3 \beta_{th} + 6\left(\frac{2\mu}{\rho R_0^2}\right)^2 \beta_{th}^2 + \right.$$

$$\left. 4\left(\frac{2\mu}{\rho R_0^2}\right)\beta_{th}^3 + 4\left(\frac{\omega^2 R_0}{2u}\right)\beta_{th}^3 + ...\right] = \frac{2^4 \gamma^4 \mu^4 R_0^4}{(3\gamma-1)^4 \sigma^4} + \frac{\gamma^4 \rho^4 R_0^{12}\beta_{th}^4}{(3\gamma-1)^4 \sigma^4} + \frac{2^5 \gamma^4 \mu^3 \rho R_0^6 \beta_{th}}{(3\gamma-1)^4 \sigma^4} +$$

$$\frac{2^3 3 \gamma^4 \mu^2 \rho^2 R_0^8 \beta_{th}^2}{(3\gamma-1)^4 \sigma^4} + \frac{2^3 \gamma^4 \mu \rho^3 R_0^{10}\beta_{th}^3}{(3\gamma-1)^4 \sigma^4} + \frac{2\gamma^4 \rho^4 \omega^2 R_0^{13}\beta_{th}^3}{(3\gamma-1)^4 \sigma^4 u} + ... \quad (28a)$$

are proportional to the radius having exponent $-4$. The only term that fulfills this condition is the second term in the expression (28a), for the thermal damping coefficient given by the expression (17c),

$$\left(\frac{\gamma\rho R_0^3 \beta}{(3\gamma-1)\sigma}\right)^4 \cong \frac{\gamma^4 \rho^4 R_0^{12}\beta_{th}^4}{(3\gamma-1)^4 \sigma^4} + ... = \frac{3^4(\gamma-1)^4 \chi^2}{2^2 \omega^6 R^4} + ... \quad (28b)$$

By replacing (28b) in the expression of force (25c), it results

$$F_{Batr}(r,R_0^6) \cong \frac{-\pi\gamma^4 \omega^2 A^4 \rho R_0^{10}}{2^3(3\gamma-1)^4 \sigma^4 r^2}\left[\frac{3^5(\gamma-1)^4 \chi^2}{2^2 \omega^2 R_0^4} + ...\right] = \frac{-3^5 \gamma^4(\gamma-1)^4 \pi A^4 \chi^2 \rho R_0^6}{2^5(3\gamma-1)^4 \sigma^4 r^2} + ... =$$

$$\frac{-m^2}{r^2}\left[\frac{3^5 \gamma^4(\gamma-1)^4 A^4 \chi^2}{2^9(3\gamma-1)^4 \pi\rho\sigma^4}\right] + ... \quad (29)$$

We note that this attractive force is independent of angular frequency and is the effect of the absorption of wave energy by the gas / vapor from the bubbles.

When $\omega \gg \omega_0$, the square bracket from the expression of the attractive force (18) can be approached

$$\left[\left(\omega^2-\omega_0^2\right)^2 + 4\beta^2\omega^2\right] = \omega^4\left[\left(1-\frac{\omega_0^2}{\omega^2}\right)^2 + \frac{4\beta^2}{\omega^2}\right] \quad (30)$$

Subtracting this new approach in (18), yields

$$F_{Batr}(r) = \frac{-2\pi A^4}{r^2 \rho^3 R_0^2 \omega^6\left[\left(1-\frac{\omega_0^2}{\omega^2}\right)^2 + \frac{4\beta^2}{\omega^2}\right]^2} = \frac{-2\pi A^4}{r^2 \rho^3 R_0^2 \omega^6\left[\left(1-\frac{p_{eff}}{\gamma\rho R_0^2 \omega^2}\right)^2 + \frac{4\beta^2}{\omega^2}\right]^2}, \quad (31)$$

Replacing the expression (13) for $\omega_0$ and the expression (7) for $\beta$, in (31), results

$$F_{Batr}(r) = \frac{-2\pi A^4}{r^2 \rho^3 R_0^2 \omega^6\left[\left(1-\frac{p_{eff}}{\gamma\rho R_0^2 \omega^2}\right)^2 + \frac{4}{\omega^2}\left(\frac{\omega^2 R_0}{2u}+\frac{2\mu}{\rho R_0^2}+\beta_{th}\right)^2\right]^2}. \quad (32)$$

This force, for any approximation of the effective pressure $p_{eff}$ and the thermal damping coefficient $\beta_{th}$, according to relations (14) and (17), does not have terms that meet the requirements of the second section.

We have shown that in the interaction between two bubbles induced by an acoustic wave there is also a gravitational acoustic force. The gravitational acoustic force exists at resonance according to formula (21) and for the low angular frequency ($\omega << \omega_0$) according to the formulas (27) and (29).

## 5. Conclusions

We have shown that under certain restrictive conditions, Bjerknes secondary forces are attractive, according to formulas (21), (27) and (29). These gravitational acoustic forces depend on the virtual mass of the oscillating bubble as the square of it.

We point out that the expression of gravitational force derived in this paper is not the final one. We can make the same statement for the electrostatic force.

We believe that one can obtain the final expressions of these forces if one investigates the interaction of $N$ identical bubbles which form a cluster. A previous analysis of this phenomenon, studied in the literature as a phenomenon specific to a cluster [7, 8], points out that the coupled oscillations of the $N$ bubbles induce a mass (as a measure of mechanical inertia and gravitational mass) for each bubble, much higher than the virtual mass. The induced mass mentioned above depends on the number of bubbles and, on the size of the cluster and of the equivalent mass of the bubble. The dependencies mentioned above are equivalent to the Mach principle in the electromagnetic universe [9, 10]. In a further paper we will try to prove the above assumption.

**References**

1. H. N. Oguz and A. Prosperetti, The natural frequency of oscillation of gas bubbles in tubes, J. Acoust. Soc. Am. 103 (6), 3301–3308, 1998.
2. R. Mettin, I. Akhatov, U. Parlitz, C. D. Ohl, and W. Lauterborn., Bjerknes forces between small cavitation bubbles in a strong acoustic field. Phys. Rev. E 56, 2924, 1997.
3. J. B. Keller, Bubble Oscillations of Large Amplitude, The Journal of the Acoustical Society of America 68 (2), 628-633, August 1980.
4. A. Prosperetti, and A. Lezzi, Bubble Dynamics in a Compressible Liquid. I. First-. Order Theory, Journal of Fluid Mechanics, 168, 457-478, 1986.
5. T. Barbat, N. Ashgriz, and C. S. Liu, Dynamics of two interacting bubbles in an acoustic field, J. Fluid Mech. 389, 137-168, 1999.
6. A. Doinikov, Acoustic radiation forces: Classical theory and recent advances, Recent Res. Devel. Acoustics, 1: 39-67, 2003.
7. I. Simaciu, Z. Borsos, Gh. Dumitrescu, G. T. Silva and T. Bărbat, Acoustic scattering-extinction cross section and the acoustic force of electrostatic type, arXiv:1711.03567, General Physics (physics. gen. ph).
8. R. Manasseh and A. Ooi, Frequencies of acoustically interacting bubbles, Bubble Science, Vol. 1, No. 1-2, 58-74, 2009.
9. J. Barbour and H. Pfister (Eds.) Mach's Principle–From Newton's Bucket to Quantum Gravity, Birkhauser, Boston (1995).